\documentclass[useAMS, usenatbib]{mn2e_letter}
\usepackage{graphicx}
\usepackage{subfigure,captcont}
\usepackage{amsmath, amsfonts, amssymb,fancyhdr}
\usepackage{natbib}
\usepackage{hyperref}

\newcommand{\kms}{\;{\rm km\; s}^{-1}}
\newcommand{\cmc}{\;{\rm cm}^{-3}}
\newcommand{\nsubhtwo}{n_{\rm H_2}}

\title{Parsec-scale magnetic fields in Arp 220}
\author[McBride et al.]
{James McBride$^{1}$\thanks{E-mail: jmcbride@astro.berkeley.edu}, 
    Timothy Robishaw$^{2}$, 
    Carl Heiles$^{1}$, 
    Geoffrey C. Bower$^{3}$, and \newauthor
    Anuj P. Sarma$^{4}$ \\
$^{1}$Department of Astronomy, University of California, Berkeley, CA 94720, USA\\
$^{2}$National Research Council Canada, Herzberg Astronomy and Astrophysics Programs, Dominion Radio Astrophysical Observatory, \\
\indent Penticton, BC V2A 6J9, Canada \\
$^{3}$Academica Sinica Institute of Astronomy and Astrophysics, 645 N. A'ohoku Place, Hilo, HI 96720, USA \\
$^{4}$Department of Physics, DePaul University, Chicago, IL 60614, USA}

\begin{document}
\maketitle
\begin{abstract}
We present the first very-long-baseline interferometry (VLBI) detections of Zeeman splitting in another galaxy.
We used Arecibo Observatory, the Green Bank Telescope, and the Very Long Baseline Array to perform dual-polarization observations of OH maser lines in the merging galaxy Arp~220. 
We measured magnetic fields of $\sim$1--5~mG associated with three roughly parsec-sized clouds in the nuclear regions of Arp~220.
Our measured magnetic fields have comparable strengths and the same direction as features at the same velocity identified in previous Zeeman observations with Arecibo alone.
The agreement between single dish and VLBI results provides critical validation of previous Zeeman splitting observations of OH megamasers that used a single large dish.
The measured magnetic field strengths indicate that magnetic energy densities are comparable to gravitational energy in OH maser clouds.
We also compare our total intensity results to previously published VLBI observations of OH megamasers in Arp~220.
We find evidence for changes in both structure and amplitude of the OH maser lines that are most easily explained by variability intrinsic to the masing region, rather than variability produced by interstellar scintillation.
Our results demonstrate the potential for using high-sensitivity VLBI to study magnetic fields on small spatial scales in extragalactic systems. \\

\end{abstract}

\section{Introduction} \label{sec:h_intro}
Magnetic fields are dynamically important in the interstellar medium (ISM) of the Milky Way \citep{Boulares1990}, and generally appear to be dynamically important within star forming galaxies similar to the Milky Way \citep[e.g.,][]{Beck2012}.
The most common method of estimating extragalactic magnetic field strengths is to use observations of synchrotron flux; if one assumes rough equality between the energy density in cosmic rays and magnetic fields, the synchrotron flux provides a measure of the magnetic field strength \citep{Beck2005}. 
If one applies this same assumption to starburst galaxies, it suggests that magnetic fields are {\em not} dynamically important in starburst galaxies \citep{Thompson2006}. 
There is evidence, however, that the assumption of equality between cosmic ray and magnetic field energy density breaks down in some environments, including starburst galaxies \citep{Thompson2006,McBride2014} and radio jets \citep{Hardcastle1998,Croston2003,McBride2014a}.

Zeeman splitting measurements provide an alternative method to probe the role of magnetic fields in some extragalactic environments, though its use is presently limited because Zeeman splitting is typically quite weak.
The only existing detections of extragalactic Zeeman splitting are in an H~I absorption line in a high velocity system toward the Perseus cluster \citep{Kazes1991,Sarma2005} and in OH maser lines in 15 starburst galaxies \citep{Robishaw2008,McBride2013} that host OH masers so powerful they are called OH megamasers \citep[OHMs;][]{Lo2005}.
While \citet{Sarma2005} used the Very Large Array (VLA), the other observations of Zeeman splitting all used single dish telescopes, and none of the Zeeman detections directly provided spatial or structural information on the fields they detected.
Moreover, OHM emission is often very complex, and overlapping features at the same velocity may blend, weakening the Zeeman splitting signal.

Dual-polarization very-long-baseline interferometry (VLBI) observations of OHMs are uniquely able to directly and accurately measure magnetic fields on small spatial scales in other galaxies.
OHMs are found almost exclusively in merging galaxies in the midst of a starburst phase \citep{Willett2011a,Willett2011}; high infrared luminosities pump the OH molecule, providing inversion for strong maser action \citep{Lockett2008}.
Therefore, measuring magnetic fields in OHMs is an opportunity to study the role of magnetic fields in star forming environments very unlike those found in the Milky Way.

\begin{figure*}
    \includegraphics[width=7in]{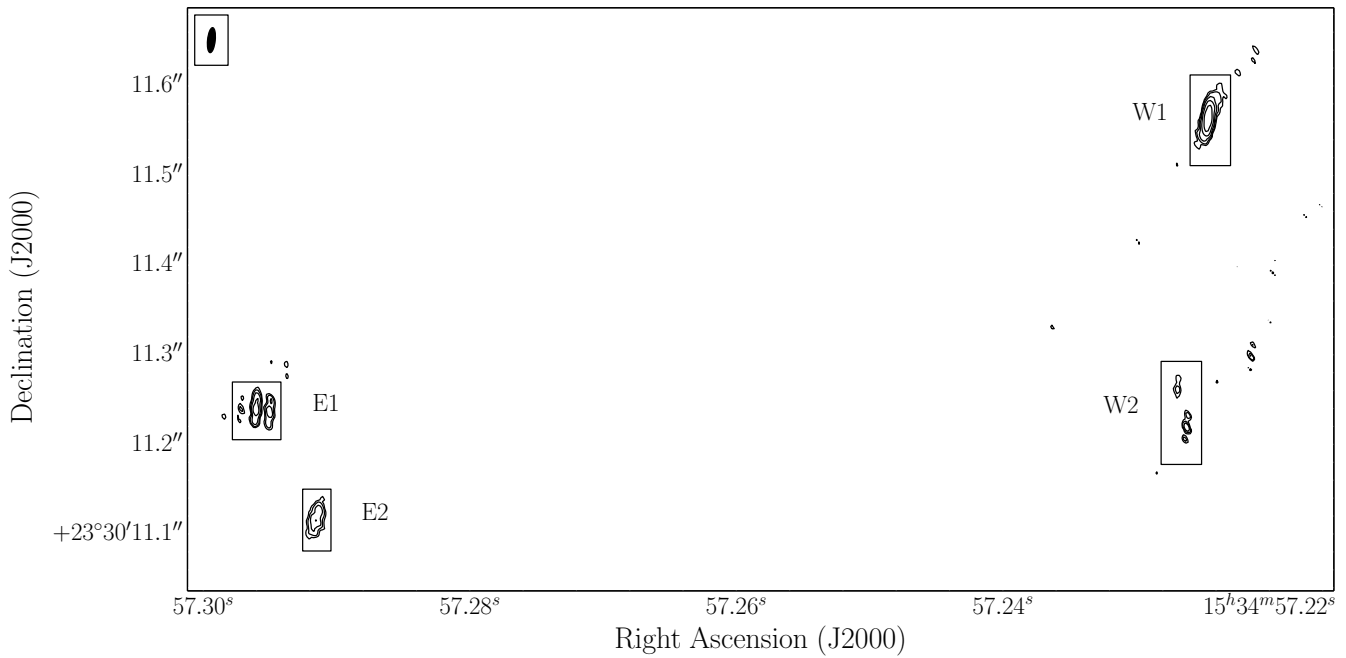}
    \caption{
        An integrated intensity map of Stokes $I$.
        In both the right-handed circularly polarized and left-handed circularly polarized data cubes, we clipped values at 3~mJy, which is 2.5 times the rms noise. 
        We then summed the two polarizations to generate the Stokes $I$ cube, and then summed over all spectral channels.
        The contours are at 3,~6,~12,~24,~48 mJy~beam$^{-1}$.
        The FWHM of the beam, which is 28 x 8.0 mas, is shown in the upper left corner of the image.
        The labels E1, E2, W1, and W2, correspond to the labels used in \citet{Lonsdale1998}.
        There are two bright features in the box labeled E1, and one dim feature.
        We refer to the more western of the bright features as E1.1, and the bright feature to the east as E1.2.
    } \label{fig:i_map}
\end{figure*}

Arp~220 is the natural first target for dual-polarization VLBI observations. 
Arp~220 is characteristic of OHMs, as it is in the midst of a merger, with two nuclei separated by roughly 1$\arcsec$, or 350~pc \citep{Graham1990,Condon1991}, that is driving a powerful starburst \citep[e.g.,][]{Smith1998,Parra2007}.
It was the first discovered OHM \citep{Baan1982}, and because of Arp~220's proximity (76~Mpc), it is the brightest known OHM (though not intrinsically the most luminous).
VLBI observations of Arp~220 have revealed multiple bright, compact maser spots with fluxes of $\sim$10--150~mJy and sizes of a few parsecs associated with both nuclei \citep{Diamond1989,Lonsdale1994a,Lonsdale1998,Rovilos2003}.
\citet[][hereafter RQH08]{Robishaw2008} used Arecibo to clearly detect Zeeman splitting associated with many narrow spectral features in the total single dish spectrum, which corresponded to magnetic fields with strengths $\sim$0.7--8~mG.
Despite the success of their observations, the lack of direct spatial information in the single dish spectrum, as well as potential line blending among features, complicates interpretation of their results.

In this paper, we have two primary aims: to test the accuracy of Zeeman splitting measurements derived from a single dish spectrum with many blended lines; and to directly probe the structure of magnetic fields in the two nuclear regions of Arp~220.
In Section \ref{sec:h_obs}, we describe our observations and data reduction. In Section \ref{sec:h_results}, we present magnetic field measurements for individual maser spots in Arp~220, and also present evidence for variability in the Arp~220 maser lines.
In Section \ref{sec:h_discussion}, we discuss the dynamic importance of the detected magnetic fields, and explore whether the observed variability is intrinsic or produced by interstellar scintillation within the Milky Way. 
Finally, we briefly recapitulate our results in Section \ref{sec:h_conclusion}.

\section{Observations \& Data Reduction} \label{sec:h_obs}
We used the combined Very Long Baseline Array (VLBA), phased VLA, 100-m Robert C. Byrd Green Bank Telescope (GBT), and the 305-m William E. Gordon Telescope at Arecibo Observatory, in a configuration called the High Sensitivity Array (HSA), to observe Arp~220 in dual-polarization mode.\footnote{The Very Long Baseline Array, the Very Large Array, and the Green Bank Telescope are operated by the National Radio Astronomy Observatory, which is a facility of the National Science Foundation operated under cooperative agreement by Associated Universities, Inc.}$^,$\footnote{The Arecibo Observatory is operated by SRI International under a cooperative agreement with the National Science Foundation (AST-1100968), and in alliance with Ana G. M\'{e}ndez-Universidad Metropolitana, and the Universities Space Research Association.}
The observations occurred on June 20, 2008 with project code BR0125.
The correlator was configured to capture a 2~MHz band with 512 spectral channels, for a frequency resolution of 3.9~kHz ($0.73 \kms$). 
We have 2.5 hours of Arecibo data, of which 2 hours was spent observing Arp~220. 
For the VLBA and GBT, we have 4 hours of data, with 3 hours spent observing Arp~220.
As a result of low level radio frequency interference (RFI) from the Iridium satellite, the VLA did not phase properly, and thus did not provide usable data.

\begin{table*}
    \centering
    \caption{Selected Zeeman Splitting Fit Results} \label{tab:zeeman}
    \begin{tabular}{rrrrrrrrr} \hline \hline
        Feature &  Figure & Gaussian   &  RQH08 Gaussian   & $S$ & $v_{\odot}$ & $\Delta v$ & $\nu$ & $B_{\parallel}$ \\
                &  &  & & (mJy) & (km s$^{-1}$) & (km s$^{-1}$) & (MHz) & (mG) \\ \hline
    W1 & \ref{fig:1}  & 2 & 9 & $                        115.0 \pm   4.1$ & $      5352.0 \pm     0.3$ & $     11.2 \pm      0.4$ & $ 1638.115$ & $               -1.5 \pm   0.4$ \\
    W1 & \ref{fig:1}  & 5 & 3 & $\phantom{8}\phantom{8}    7.8 \pm   1.8$ & $      5455.8 \pm     0.7$ & $      4.0 \pm      1.0$ & $ 1637.558$ & $\phantom{-}     7.0 \pm   2.5$ \\ 
  E1.1 & \ref{fig:2}   & 1 & 4 & $              31.7 \pm   1.9$ & $      5426.3 \pm     0.2$ & $      7.9 \pm      0.5$ & $ 1637.716$ & $    -5.2 \pm   0.9$ \\
  E1.2 & \ref{fig:3}  & 4 & 4 & $              48.3 \pm   2.2$ & $      5428.4 \pm     0.2$ & $      9.8 \pm      0.7$ & $ 1637.705$ & $\phantom{ }    -3.0 \pm   0.6$ \\ 
    E2 & \ref{fig:4}  & 4 & --- & $              36.6 \pm   1.2$ & $      5403.0 \pm     0.1$ & $     16.7 \pm      0.4$ & $ 1637.841$ & $           \phantom{ }    -3.4 \pm   1.2$ \\ \hline
    \end{tabular}
    \vskip 0.25em 
    \raggedright
    {{\bf Note: } The column labeled ``Gaussian'' specifies the component of the fit, as labeled in Figure \ref{fig:iv}.
        For instance, the feature E1.2 is shown in Figure \ref{fig:3}, and we fit Gaussian 4 to have a magnetic field $B_\parallel = -3.0 \pm 0.6$~mG.
        Though the feature for E2 is also labeled Gaussian 4, it corresponds to a different feature, shown in Figure \ref{fig:4}, which has $B_\parallel = -3.4 \pm 1.2$.
        The column labeled ``RQH08 Gaussian'' shows the label used in Table 8 and Figure 8 of \citet{Robishaw2008} that appears to match the feature we see in the VLBI data.
        The relevant information from Table 8 in RQH08 is reproduced here in Table \ref{tab:r08}.
The reported errors for the fits are purely statistical.}
\end{table*}

\begin{table*}
    \centering
    \caption{Relevant Zeeman Splitting Results from \citet{Robishaw2008}} \label{tab:r08}
    \begin{tabular}{rrrrrr} \hline \hline
      Gaussian & $S$ & $v_{\odot}$ & $\Delta v_{\odot}$ & $\nu$ & $B_{\parallel}$ \\
         & (mJy) & (km s$^{-1}$) & (km s$^{-1}$) & (MHz) & (mG) \\ \hline
    3 & $              10.1 \pm   1.0$ & $      5452.9 \pm     0.2$ & $      4.5 \pm      0.5$ & $ 1637.574$ & $    7.77 \pm   0.76$ \\
    4 & $              90.5 \pm   0.8$ & $      5425.6 \pm     0.1$ & $      11.0 \pm      0.1$ & $ 1637.720$ & $    -2.78 \pm   0.13$ \\
    9 & $              386.2 \pm   5.9$ & $      5351.2 \pm     0.1$ & $      16.5 \pm      0.2$ & $ 1638.119$ & $    -0.76 \pm   0.06$ \\ \hline
    \end{tabular}
    \vskip 0.25em
    \raggedright
    {{\bf Note:} Fits from RQH08 for which we see corresponding features in the HSA data. The Gaussian number here refers to the labels used in Table 8 and Figure 8 of RQH08.}
\end{table*}

We used the Astronomical Image Processing System \citep[AIPS;][]{Greisen2003} for all phase and amplitude calibration of the $uv$ data, including many of the VLBI~specific tasks described in Appendix C of the AIPS Cookbook. 
We followed the general procedure for reducing polarized VLBI data provided in the Appendix of \citet{Fish2005}, though a few steps required for full-polarization data did not apply to our dual-polarization data.
We used 3C~286 as a bandpass calibrator and used J1532+2344\footnote{Flux information for J1532+2344 is available at \href{http://www.vlba.nrao.edu/astro/calib/vlbaCalib.txt}{http://www.vlba.nrao.edu/astro/calib/vlbaCalib.txt}} to calculate instrumental phases and delays. 
We did not observe J1532+2344 frequently enough to solve for time varying phases, so we performed phase calibration using self-calibration of the brightest maser spot in Arp~220, and then applied that solution to the other spectral channels. 
Neither of the calibration sources we observed are ideal flux calibrators; 3C~286 has resolved structure at our resolution, and J1532+2344 is dim and does not have a reported flux at our observed frequency.
Thus, for amplitude calibration, we used the recorded system temperatures for the telescopes in the array.
Comparing our measured fluxes for 3C~286 with the results from \citet{Kus1988}, we estimate $\sim$30\% systematic uncertainty in our amplitude calibration.\footnote{Magnetic field strengths derived from Zeeman splitting are independent of absolute amplitude calibration, as the Zeeman signal depends upon the relative amplitudes of Stokes $I$ and $V$.}

The system temperatures for the telescopes that contributed data ranged from 20--30~K. 
The OH lines at their brightest have fluxes of $\sim$200~mJy.\footnote{As in RQH08 and \citet{McBride2013}, we use the classical definition of Stokes I, which is the sum of the left- and right-circularly-polarized emission. Some other work uses the average instead.}
For gains ranging from $\sim$0.1--10~K/Jy, the lines themselves thus make a minimal contribution to the rms noise, meaning any Stokes~$V$ signal we detect does not depend sensitively on artifacts from finding the difference of two relatively bright signals \citep{Liszt2002a}. 

After applying all solutions, we used the AIPS task IMAGR to separately generate data cubes for the right-handed circular polarization ($RHCP$) and left-handed circular polarization ($LHCP$), and saved the images in FITS files for analysis elsewhere.
For both the $RHCP$ data cube and $LHCP$ data cube, the rms error was 1.2~mJy~beam$^{-1}$~channel$^{-1}$. 
For the Stokes $I$ data cube, which is the sum of the $RHCP$ and $LHCP$, the rms error was 1.7~mJy~beam$^{-1}$~channel$^{-1}$. 
The beam had a full-width at half maximum (FWHM) of $28 \times 8.0$~mas. 
For a distance of 76~Mpc to Arp~220, the physical resolution is $10 \times 2.9$~pc.

\section{Results} \label{sec:h_results}

We identified prominent maser spots for Zeeman splitting analysis using a map of the total Stokes~$I$ emission. 
To produce the map, which is shown in Figure \ref{fig:i_map}, we clipped the Stokes~$I$ data cube at 3 times the rms noise, and then flattened the image over all spectral channels. 
The map is qualitatively similar to that presented in \citet[][hereafter, R03]{Rovilos2003}; we will compare our Stokes~$I$ results to theirs in Section \ref{sec:comp}.
\citet{Lonsdale1998} discussed the four brightest regions of emission, and labeled them W1, W2, E1, and E2, where ``W'' refers to the western nucleus, ``E'' to the eastern nucleus, ``1'' to the northern component, and ``2'' to the southern component.
We will follow their nomenclature, and labelled the spots accordingly in Figure \ref{fig:i_map}.
We also use two additional labels: E1.1 refers to the western of the two bright features in E1, and E1.2 refers to the eastern of the two bright features in E1 (there is a third dim feature in E1 that we do not discuss).

We then generated $RHCP$ and $LHCP$ spectra from their respective {\em unclipped} data cubes.
For each distinct maser spot, we selected the brightest pixel, and summed neighboring pixels in a region corresponding to 95\% of the beam power. 
To account for the imperfect amplitude calibration of the $RHCP$ and $LHCP$ components, we solved for a single leakage solution for all maser spots.
With the leakage corrected polarizations, we then took $I = RHCP + LHCP$ and $V = RHCP - LHCP$.

\begin{figure*}
    \centering
    \subfigure[Stokes $I$ and $V$ for the OH maser spot at \newline W1 (15$^h$34$^m$57$^s$.225, +23$^\circ$30$\arcmin$11.562$\arcsec$).]{
        \label{fig:1}
        \includegraphics[width=3.0in] {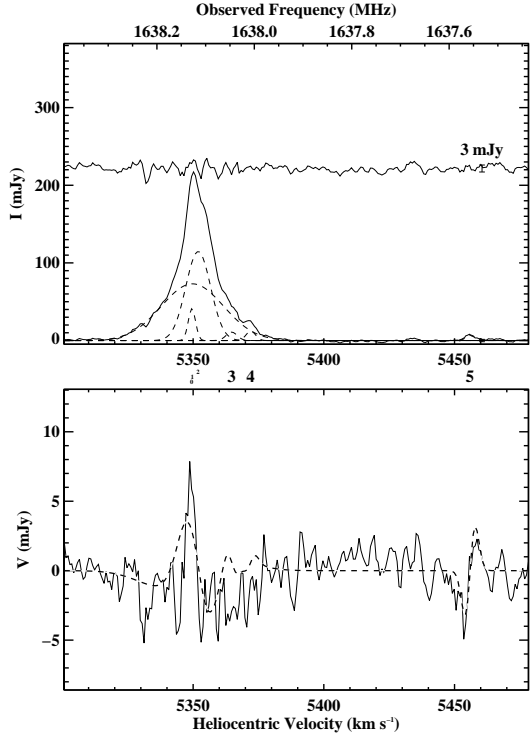}
    }
    \subfigure[Stokes $I$ and $V$ for the OH maser spot at \newline E1.1 (15$^h$34$^m$57$^s$.295, +23$^\circ$30$\arcmin$11.235$\arcsec$).]{
        \label{fig:2}
        \includegraphics[width=3.0in] {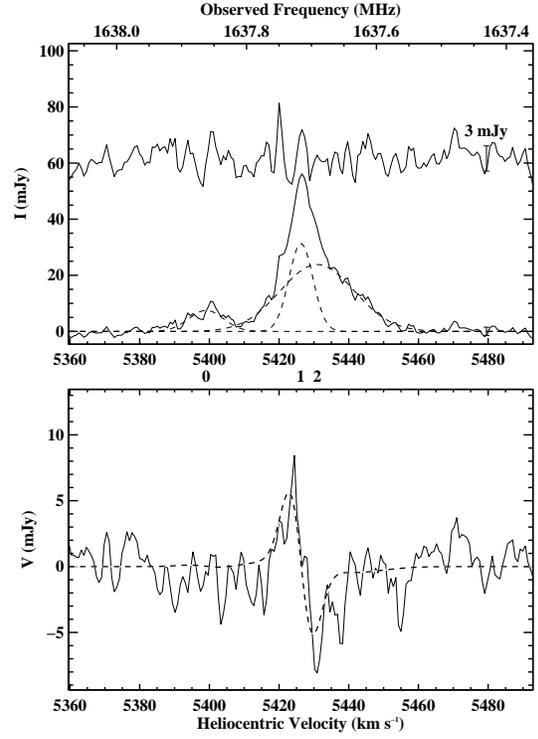}
    } \\
    \subfigure[Stokes $I$ and $V$ for the OH maser spot at \newline E1.2 (15$^h$34$^m$57$^s$.296, +23$^\circ$30$\arcmin$11.240$\arcsec$).]{
        \label{fig:3}
        \includegraphics[width=3.0in] {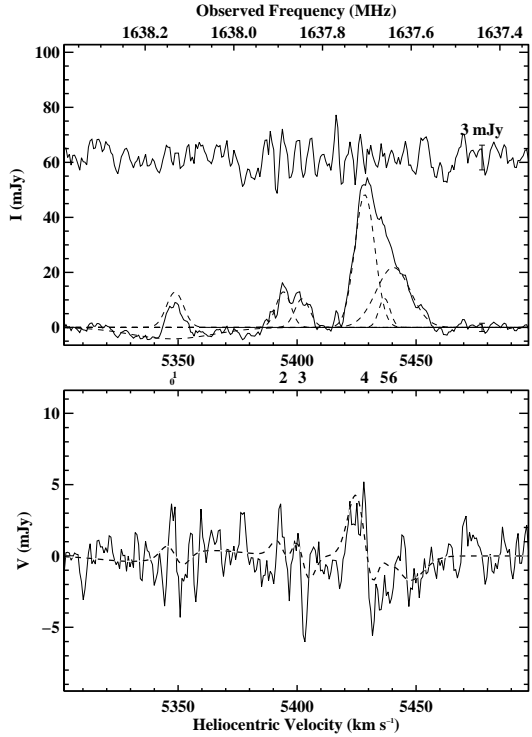}
    }
    \subfigure[Stokes $I$ and $V$ for the OH maser spot at \newline E2 (15$^h$34$^m$57$^s$.292, +23$^\circ$30$\arcmin$11.115$\arcsec$).]{
        \label{fig:4}
        \includegraphics[width=3.0in] {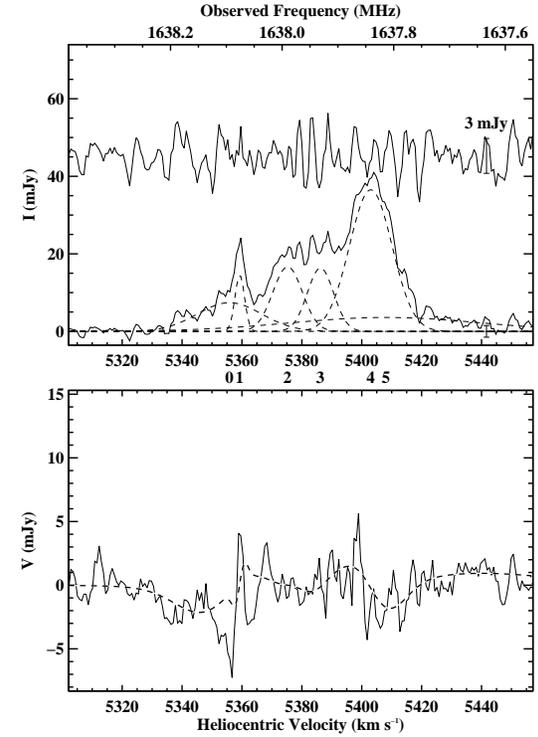}
    }
    
    \caption{
        Within each sub-figure, Stokes $I$ (top panel) and $V$ (bottom panel) spectra are shown for different maser spots. 
        Stokes $I$ is defined here as the sum of the cross-hand polarizations.
        In the top panel, the Stokes~$I$ spectra, smoothed by three channels, are represented with a solid line.
        The dashed line represents individual Gaussian fits; the corresponding numbers for each Gaussian are shown between the top and bottom panels.
        The thin solid line, offset from zero, shows the magnified residuals to the fitting.
        In the bottom panel, the solid line indicates the Stokes~$V$ spectrum smoothed by three channels, while the dashed line shows the total fit from all input Gaussians.
    } \label{fig:iv}
\end{figure*}

\subsection{Magnetic field derivation}

For OHM lines, the line splitting is generally smaller than the intrinsic line width, in which case the amplitudes of features in the Stokes $V$ spectrum are proportional to the derivative of the Stokes $I$ spectrum and the line-of-sight magnetic field strength, $B_\parallel$, with
\begin{equation}
    V = \left(\frac{\nu}{\nu_0}\right)\left(\frac{dI}{d\nu}\right)b B_\parallel + aI. \label{eq:narrow_split}
\end{equation}
Here, $b$ is the splitting coefficient, which sets the strength of the frequency splitting in the presence of a magnetic field \citep{Heiles1993}. 
For the 1667~MHz line in OH, it is 1.96~Hz~$\mu$G$^{-1}$.
The coefficient $a$ accounts for leakage, so we set $a = 0$ as a result of already applying a leakage solution.  

Many of the spectra associated with individual spots are quite complex, despite corresponding to small spatial scales.
For this reason, a single $B_\parallel$ solution that fits the total Stokes~$V$ profile to the derivative of the Stokes~$I$ profile generally did a very poor job of reproducing the observed Stokes~$V$ profile. 
The reported fits had large errors relative to their magnitudes, and visually, the fitted Stokes~$V$ did not capture the Zeeman signal associated with the brightest parts of the Stokes~$I$ spectrum.

Instead of fitting a total magnetic field strength to each Stokes~$I$ and $V$ pair, we followed the procedure used in the single dish observations of RQH08 and \citet{McBride2013} to simultaneously fit multiple features to each spectrum, as follows. 
We guess the minimum number of Gaussian features that adequately describe the structure in the Stokes $I$ spectrum, and then simultaneously fit all of the guessed parameters to the Stokes~$I$ spectrum.
Then, the Gaussian fits to a Stokes $I$ profile are used as inputs to fit the Stokes $V$ profile for that spot. 
Each Gaussian component is permitted to take a different magnetic field strength, but is fixed in frequency and width to be the same as its corresponding feature in Stokes $I$. 
While this approach means there is some degeneracy in solutions, we find that the derived $B_\parallel$ values we report are fairly insensitive to the exact input guesses we make. 
Somewhat different combinations of Gaussian parameters that still adequately fit the Stokes~$I$ and $V$ profiles yield comparable estimates of the magnetic field strength associated with the brightest features in the Stokes $I$ and $V$ spectra.

\subsection{Magnetic field detections} 

We summarize our most notable Zeeman splitting results in Table \ref{tab:zeeman}, which includes three detections and two intriguing fits that fall short of the threshold for detection. 
We decided not to report every fitted Gaussian for all features, a total of $\sim$$40$ components.
For the maser spots discussed in the text (W1, E1.1, E1.2, and E2), the upper limits for other components are all much larger than the values for the components in Table \ref{tab:zeeman}.
The spectra from which these fits were derived are shown in Figure \ref{fig:iv}.
The data plotted in Figure \ref{fig:iv} are smoothed by three channels, but all fits that we report are to the unsmoothed data.
The fits listed in the first, third, and fourth rows are the only three fits for which the fitted magnetic field strength was larger than 3 times the error associated with the fit.
The detections are not a result of particularly strong fields; the three detected fields are found among the four brightest features in the Arp~220 spectrum at any spatial location.
For ease of comparison with the results of RQH08, Table \ref{tab:r08} shows selected Zeeman splitting results from RQH08 for which there are corresponding features in the HSA data.

The third and fourth rows in Table \ref{tab:zeeman} show fits to the two bright features along the northern ridge in the eastern nucleus (E1.1 and E1.2).
The two features are at very similar velocities, and have magnetic fields oriented in the same direction, though with different strengths.
These features combined appear to correspond to Gaussian 4 from RQH08, which we show here in the second row of Table \ref{tab:r08}.
RQH08 fit a feature with $v = 5425.6 \kms$, $\Delta v = 11 \kms$, and $S = 90.5$~mJy (the flux they report, 0.90~mJy, is a typo) that has $B_\parallel = -2.78 \pm 0.13$~mG. 
Their $B_\parallel$ is in good agreement with one of the two features we see, and somewhat lower than the other. 
If we take the weighted mean of the two features we measure (using the variance in the magnetic field measurement), this gives $B_\parallel = -3.7 \pm 0.5$~mG, which is not significantly different from the value reported in RQH08.

For the brightest individual feature, in W1, we find $B_\parallel = -1.5 \pm 0.4$~mG.
By comparison, RQH08 found $B_\parallel = -0.76 \pm 0.06$~mG for the corresponding feature in their spectrum (Gaussian 9, in the third row of Table \ref{tab:r08}). 
While these values are not in perfect agreement, they are of comparable magnitude, and have the same orientation.
Along with the results for E1, this result suggests that while line blending may reduce the value of fitted fields in single dish Zeeman observations of OHMs, the effect of blending is, at most, moderate. 
Thus the previous interpretation of single dish results appears reasonable, though a more extensive comparison between single dish and VLBI results would still be valuable.

\begin{figure*}
    \centering
        \subfigure[Spectrum 1 from Figure 5 of R03 (E1.1), located at \newline 
        15$^h$34$^m$57$^s$.295 +23$^\circ$30$\arcmin$11.235$\arcsec$.]{
     \label{fig:rovi1}
    \includegraphics[width=3.2in]{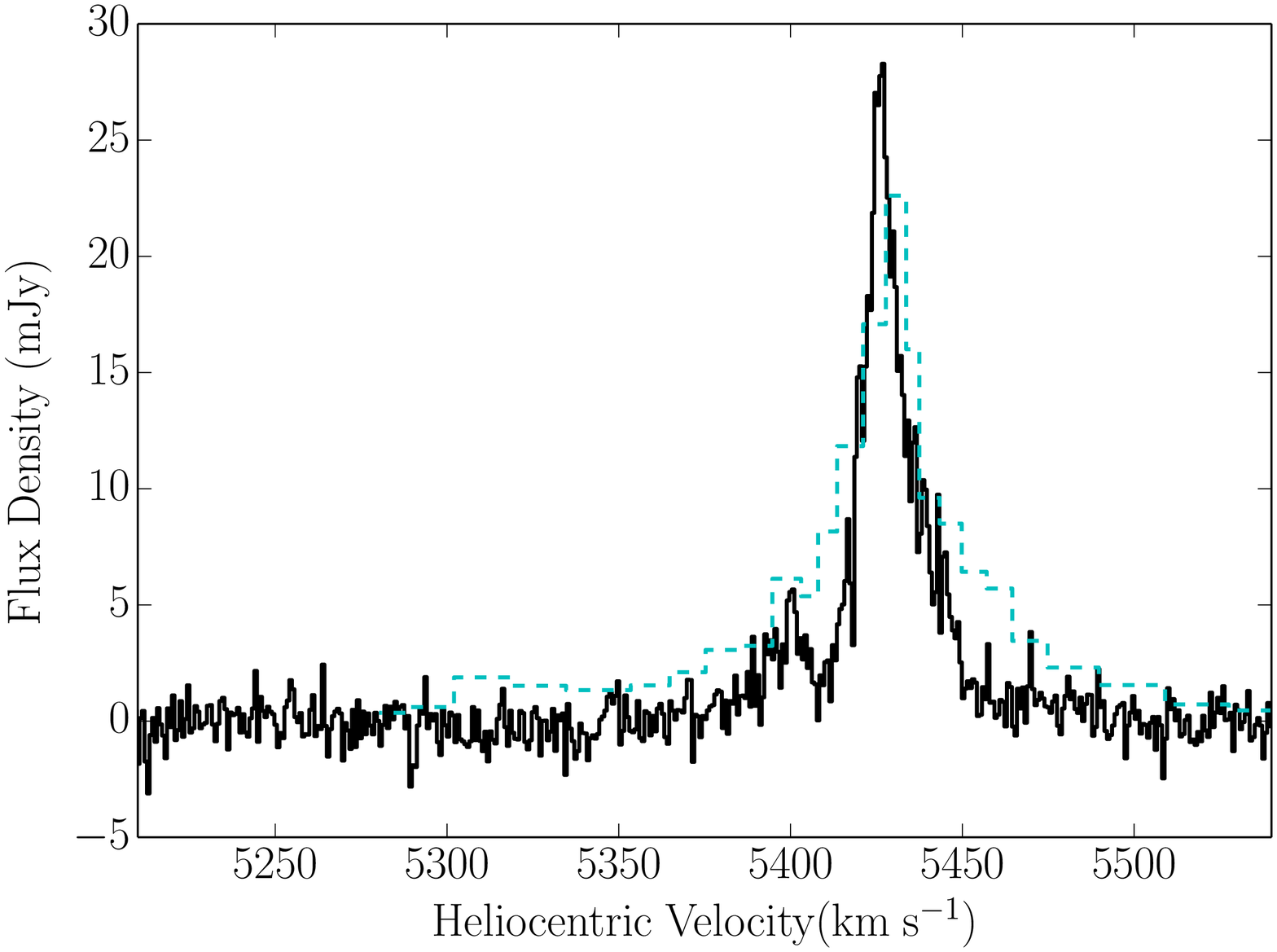}
    }
    \subfigure[Spectrum 2 from Figure 5 of R03 (E1.2), located at \newline
    15$^h$34$^m$57$^s$.296 +23$^\circ$30$\arcmin$11.240$\arcsec$.]{

     \label{fig:rovi2}
    \includegraphics[width=3.2in]{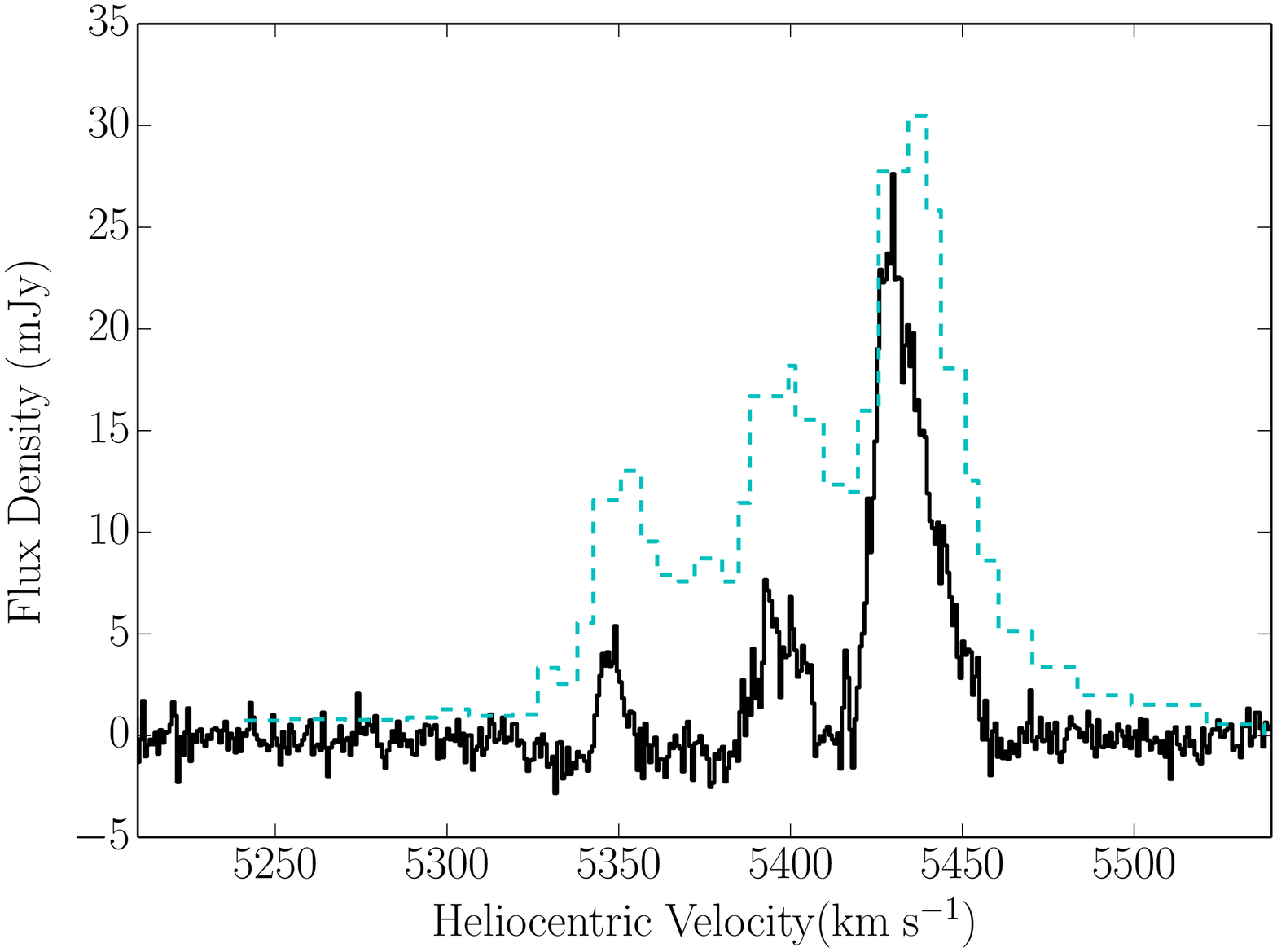}
    } \\
    \subfigure[Spectrum 5 from Figure 5 of R03 (E2), located at \newline
    15$^h$34$^m$57$^s$.292 +23$^\circ$30$\arcmin$11.115$\arcsec$.]{
     \label{fig:rovi5}
    \includegraphics[width=3.2in]{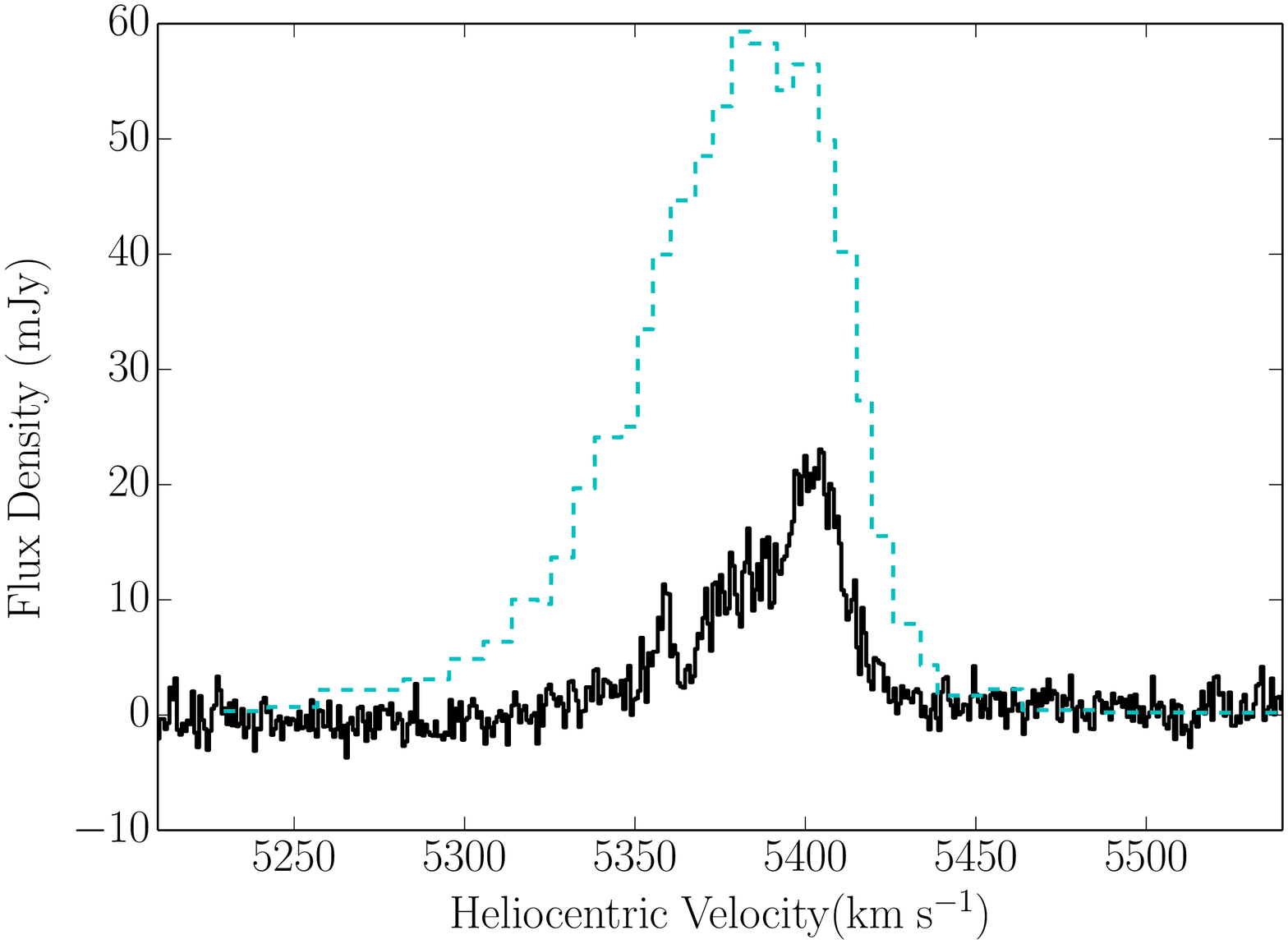}
    }
    \subfigure[Spectrum 8 from Figure 5 of R03 (W2), located at \newline
    15$^h$34$^m$57$^s$.227 +23$^\circ$30$\arcmin$11.260$\arcsec$.]{
     \label{fig:rovi8}
    \includegraphics[width=3.2in]{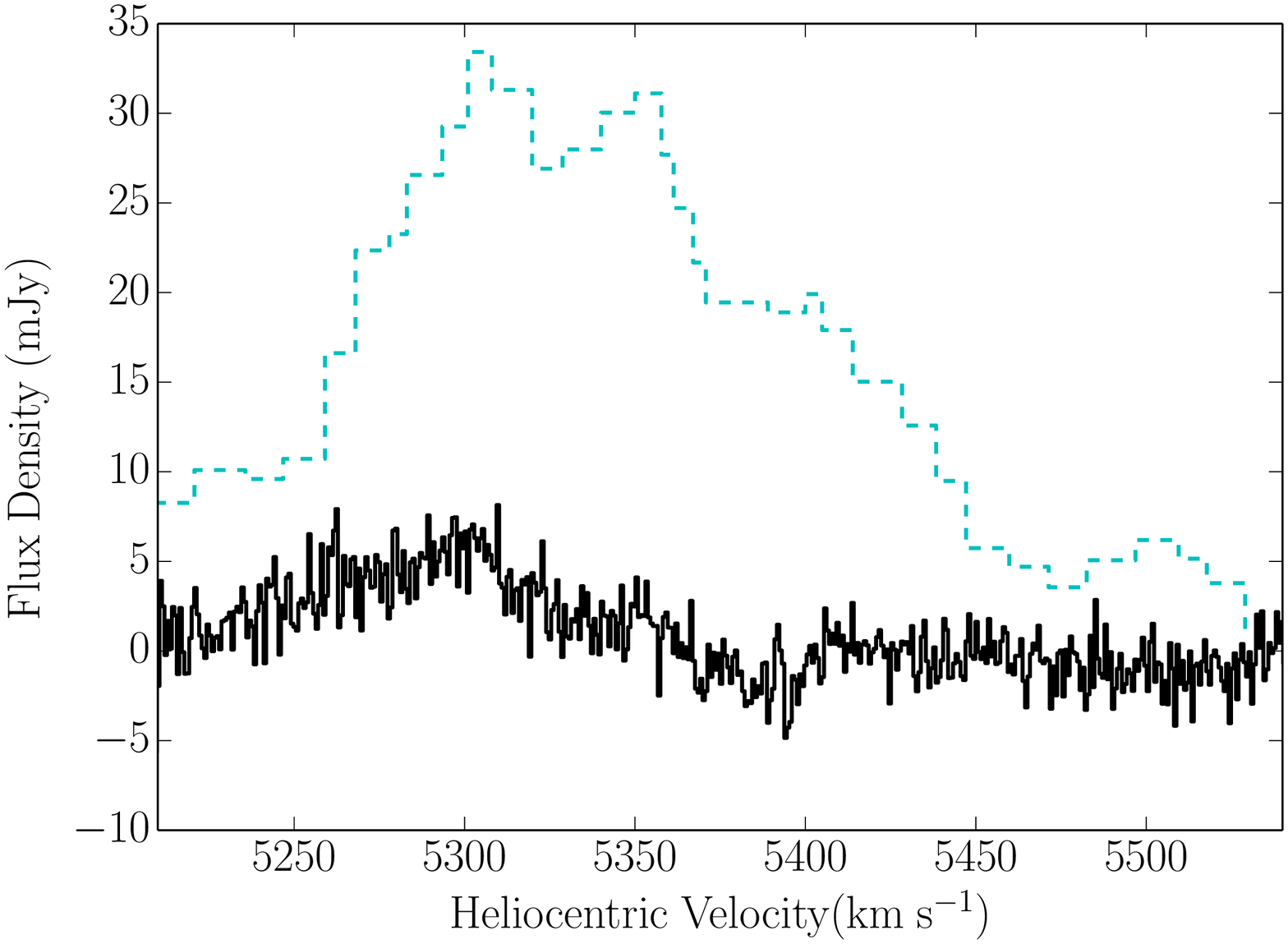}
    }
    \caption{In each subfigure, we compare our spectra (solid black) at different locations to the spectra from R03 (dashed cyan) at the same location.
We assume that they used the more common definition of Stokes $I$, and thus halve our fluxes in this figure for the purpose of comparison.
        In making these comparisons, we summed flux in boxes that replicate those used in R03 as closely as possible. 
    Experimenting with different boxes indicates that our results are not sensitive to our choice of boxes.
A larger source of error is that R03 were unable to provide spectra; 
what we show here was digitized from Figure 5 in their paper.} \label{fig:comps}
\end{figure*}

The second row in Table \ref{tab:zeeman} shows a comparatively modest feature, also in W1, that has an amplitude of 7.8~mJy.
While its fitted magnetic field is not statistically significant, the feature appears to match the RQH08 fit shown in the first row of Table \ref{tab:r08}, for which they reported a magnetic field of 7.77 $\pm$ 0.76~mG. 
The two lines have very similar amplitudes (7.8 $\pm$ 1.8~mJy, 10.1 $\pm$ 1.0~mJy) and velocity widths ($4.0 \pm 1.0 \kms$, $4.5 \pm 0.5 \kms$). 
Their velocity centers are statistically significantly different, however, with centers of $5455.8 \pm 0.7 \kms$ and $5452.9 \pm 0.2 \kms$.
Given the similarities in the other features, however, and the lack of any other compelling match for Gaussian 3 from RQH08 in the HSA data, we consider it very likely that the features are physically the same, and that the masing cloud that produces this narrow line accelerated in the time between the observations of RQH08 in 2006 and the HSA observations discussed here, from 2008.

If the identification of the two features is correct, it emphasizes the complexity of structure along the line-of-sight in a few parsec-sized cloud. 
The two features that we fit for W1, in Table \ref{tab:zeeman}, are at the same location, yet are roughly $100 \kms$ apart in velocity, and have oppositely oriented magnetic fields whose magnitudes differ by a factor of $\sim$4. 
It also confirms that the magnetic fields measured from maser lines correspond to the fields in the masing clouds themselves, rather than probing the field on larger scales in the ISM.

The fifth row in Table \ref{tab:zeeman} shows a feature from the spectrum of the bright maser spot in E2, though its corresponding $B_\parallel$ fit is not a detection.
Though it is bright, this feature is not easily identifiable in the single dish spectrum; there is no clear analog in the fits to the single dish spectrum in RQH08. 
Thus, while single dish observations are capable of providing significant information on magnetic fields in other galaxies, VLBI observations will be vital to fully understanding the detailed structure of magnetic fields. 

\subsection{Evidence for variability} \label{sec:comp}
\citet{Lonsdale1998} performed VLBI observations of Arp~220 on November 13, 1994. 
Their sensitivity was similar to ours, but as result of their superior $uv$ coverage, their observations are of higher spatial resolution.
R03 then presented more detailed maps and spectra of Arp~220 made from the same 1994 data set.
Given the evidence for time variability in OHMs \citep{Darling2002b}, including weak variation in the single dish spectrum of Arp~220 \citep{McBride2013} and in VLBI observations of Arp~220 \citep{Lonsdale2007}, we evaluate evidence for variations between the VLBI observations.

The location of maser spots is unchanged between our observations and those of R03. 
We detect emission from each spot that was boxed in Figure 5 of R03, and do not see significant ($>$2~mJy) emission from any spot that did not appear in the R03 map.
We do, however, see evidence for variations in the amplitude and structure of spectral lines.
Figure \ref{fig:comps} shows four spectra from R03 compared to our spectra at the same locations.
\citet{Lonsdale1998} and R03 do not explicitly state their Stokes~$I$ definition, but based on their discussion of their data relative to other published work, including discussion in \citet{Lonsdale1994a}, they appear to use the more common definition of the average of the two cross-hand polarizations; we assume that they used the more common Stokes~$I$ definition going forward. 
With this assumption, our amplitudes are lower for most major features that appear in the VLBI spectra.
We estimate that our amplitude scaling is $\sim$40\% lower than that in R03, based on minimizing the total difference between all features in our spectra and theirs and assuming a constant total flux in the VLBI components. 
While this scaling is large, it is within reason, given moderate systematic uncertainty in our amplitudes as a result of using the system temperature for calibration, and the same systematic uncertainty in their observations.
We do not apply this scaling in comparisons we make below, but even if we did, it would not change any of the results that follow.

Despite uncertainty in absolute amplitude calibration, the relative amplitude changes provide strong evidence of variability.
R03 measured the three brightest features in the eastern nucleus to have peak flux densities of $\sim$22, $\sim$30, and $\sim$58~mJy. These respectively correspond to spectra 1 and 2 (along the ridge in E1) and spectrum 5 (E2) in Figure 5 of their paper, shown here in Figures \ref{fig:rovi1}, \ref{fig:rovi2}, and \ref{fig:rovi5}.
The relative brightness of the features has dramatically changed, as we now measure peak flux densities of 28, 28, and 22~mJy (using their likely definition of Stokes~$I$). 
Changes in line structure are also apparent. 
In Figure \ref{fig:rovi2}, their spectrum has three distinct peaks at velocities $\sim$5350, 5400, 5430~$\kms$, with moderate flux in between the three peaks, whereas in our spectrum the flux between the peaks falls all the way to zero. 
Moreover, the relative amplitude of these features changes, with R03 measuring peak flux densities with ratios $\sim$3:1.5:1, compared to $\sim$5:1.25:1.
The most dramatic change in flux is shown in Figure \ref{fig:rovi8}, (spectrum 8 in Figure 5 of R03), which they measured to have a peak flux density of $\sim$30~mJy, and in which we measure a peak flux density of $\sim$4~mJy. 
When they observed this feature, it spanned $\sim$5200--5450~$\kms$, while we measured positive flux from $\sim$5230--5370~$\kms$, and a separate, though difficult to see, feature at $\sim$5400--5440~$\kms$.

\begin{figure}
    \centering
    \includegraphics[width=3.5in]{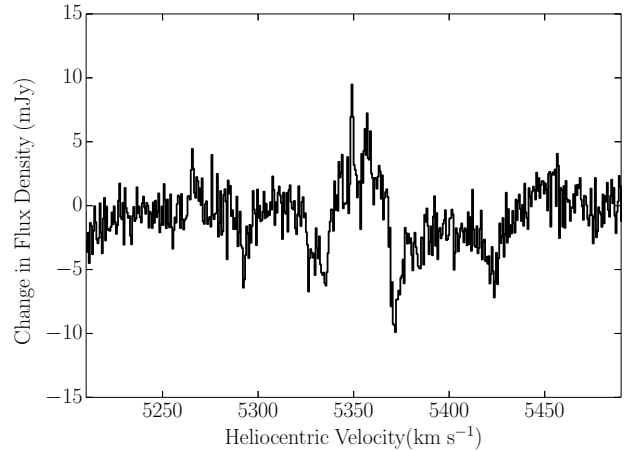}
    \caption{Variation in the amplitude of OH lines for Arp~220 between 2006 and 2008 (positive values correspond to features that got brighter from 2006 to 2008).
        The two observations were made with Arecibo, and originally presented in RQH08 and \citet{McBride2013}; the single dish evidence for variability was also discussed in Section 5.1.2 of \citet{McBride2013}.
        We present a modified version of their plot for ease of comparison with the VLBI data.
        As in Figure \ref{fig:comps}, here we use the more typical Stokes~$I$ definition.
    } \label{fig:ao_comp}
\end{figure}

While errors in amplitude calibration may account for some of the differences between spectra shown in Figure \ref{fig:comps}, it cannot explain the significant relative variations of spectral features between the 1994 data presented in R03 and the data from 2008 that we present here.
The observed variation cannot be explained by phase calibration errors either.
If phase errors were significant, they would produce spurious maser spots, but the maser spots we observe are in the same locations as those R03 observed. 
Finally, resolution differences are also unlikely to explain the observed changes. 
We had a similar, though somewhat larger, synthesized beam than R03 had.

Our result is also broadly consistent with the previous evidence for variability in the narrow OH maser lines of Arp 220.
Over the course of only two years, from 2006 to 2008, the brightest features in the Arecibo spectrum of Arp~220 experienced amplitude changes of 5--10~mJy \citep{McBride2013}. 
In Figure \ref{fig:ao_comp}, we show a modified version of Figure 3(d) from \citet{McBride2013}, which shows the variability they observed. 
The most dramatic change in the VLBI spectra occurred in E2, shown in Figure \ref{fig:rovi5}, which dimmed by $\sim$50~mJy at velocities $\sim$$5370$--$5380 \kms$. 
At the same velocity, \citet{McBride2013} observed a $\sim$10~mJy decline from 2006 to 2008.
The VLBI observations of \citet{Lonsdale2007} showed ``strong variability'' in a few spectral features that were observed 15 months apart. 

While the sizes and luminosities of OH megamaser spots exceed those of galactic OH masers, variability in galactic OH masers provides another point of comparison.
There is evidence for strong variability on long timescales in a moderate fraction of galactic OH masers, and only a small fraction of galactic OH masers are highly stable over long periods of time \citep{Caswell2013,Caswell2014}.
The variability we present here, between data acquired $\sim$14 years apart, is dramatic, but is not out of line with previous work on variability of OH masers. 

\section{Discussion} \label{sec:h_discussion}
\subsection{Magnetic fields in Arp 220}
With dual-polarization VLBI observations, we hoped to be able to directly study the structure of magnetic fields in Arp 220.
The significant decrease in flux in the southern masing components around both the eastern and western nuclei, and our lower than hoped for sensitivity as a result of RFI, make discussion of structure around the nuclei difficult.
That the two magnetic fields associated with E1.1 and E1.2 have the same direction and comparable amplitudes suggests at least the possibility of larger scale magnetic structure. 
On the other hand, the two components in W1 that we discussed have oppositely oriented fields. 
Though there may be structure in W1 along the line of sight, it may also indicate a disordered field.
Without additional detections, we will not speculate further. 

\begin{table}
    \caption{Dynamic importance of detected fields} \label{tab:dynamic}
    \begin{tabular}{rrrrrr} \hline \hline
    feature & $\Delta v$ & $B_{\parallel}$ &  $R$ & $\mathcal{M} / \left|\mathcal{W}\right|$ & $\mathcal{M} / \mathcal{T}$ \\
            & (km s$^{-1}$) & (mG) & (pc) & & \\ \hline
(E1.1) & $7.9 \pm 0.5$ & $-5.18 \pm 0.88$ & 1 & 1.4 & 3.5 \\
(E1.2) & $9.8 \pm 0.7$ & $-2.96 \pm 0.61$ & 1 & 0.5 & 0.8 \\ 
(W1)  & $11.1 \pm 0.4$ & $-1.74 \pm 0.36$ & 1 & 0.2 & 0.2 \\ \hline
\end{tabular}
    \vskip 0.25em
{{\bf Note: } We use the results of \citet{Lonsdale1998} for size estimates, as their spatial resolution was better than ours. They note that the full extent of W1 is 25 x 1 pc. Their position velocity diagram shows that the $5350 \kms$ feature peaks in the central few milliarcseconds, corresponding to a physical size of roughly a parsec.
    For the features in E1, \citet{Lonsdale1998} say that they are barely resolved with sizes 1--2~pc. 
Thus we take 1~pc as a characteristic radius for all spots.
The ratios $\mathcal{M} / \left|\mathcal{W}\right|$ and $\mathcal{M} / \mathcal{T}$ are defined in Equations \ref{eq:mg} and \ref{eq:mt}, respectively.}
\end{table}

We can, however, address the dynamical importance of the magnetic fields we detect.
As in \citet{McBride2013}, we follow the \citet{Stahler2005} application of the virial theorem to a spherical cloud. 
Assuming the cloud's mass is dominated by molecular hydrogen, the ratio of the magnetic energy density, $\mathcal{M}$, to the self gravitational energy, $\mathcal{W}$, is
\begin{equation}
    \frac{\mathcal{M}}{\left|\mathcal{W}\right|} \simeq 0.5 \left(\frac{B}{3 {\rm \;mG}}\right)^2 \left(\frac{R}{1 {\rm \; pc}}\right)^{-2} \left(\frac{\nsubhtwo}{10^5 \cmc}\right)^{-2}. \label{eq:mg}
    \end{equation}

Likewise, the ratio of the magnetic energy density to the turbulent energy density, $\mathcal{T}$, is 
\begin{align}
    \frac{\mathcal{M}}{\mathcal{T}} & \simeq 0.8 \left(\frac{B}{3 {\rm \;mG}}\right)^2  \times \nonumber \\ & \left(\frac{\Delta v}{10 {\rm \; \kms}}\right)^{-2} \left(\frac{\nsubhtwo}{10^5 \cmc}\right)^{-1}. \label{eq:mt}
\end{align}

The $\Delta v$ here is an internal bulk velocity dispersion within the cloud. 
We use our measured linewidths as an estimate of $\Delta v$, though the two values are not precisely the same. 
For the number density, $\nsubhtwo$, we conservatively take $10^5 \cmc$. 
At densities much above $10^5 \cmc$, masing is shut down as OH molecules thermalize \citep[M. Elitzur 2013, private communication;][]{Elitzur1992book,Parra2005}.
As a point of reference, our assumed density is a factor of 30 larger than the value used by \citet{Parra2005} to model the combined compact and diffuse OHM emission in III~Zw~35, and a factor of 10 larger than the model density used by \citet{Lockett2008}.

Altogether, this allows us to estimate the relative energy densities of magnetic fields, gravity, and turbulent motions in Table \ref{tab:dynamic}.
With our measurements, and assumption about $\nsubhtwo$, magnetic fields in Arp~220 have energy density of order the gravitational energy and turbulent energy. 
While a moderately higher number density would drive the value down, thermalization does not permit the density to increase significantly.
Moreover, our measured $B_\parallel$ represents a lower limit to the total magnetic field strength, $B$. 
For a few different reasonable distributions of magnetic field strengths in a region, the mean or median $B_\parallel \simeq 0.5 B$ \citep{Heiles2005,Crutcher2010}.
Thus the evidence supports magnetic fields being dynamically important in masing clouds in Arp~220. 

\subsection{Source of observed variability}
In the first, and most detailed, examination of variability in OHMs, \citet{Darling2002b} considered whether the variability they observed was produced intrinsically or via interstellar scintillation.
For the moderate variability they observed on short timescales, they concluded that the evidence was ambiguous, but interstellar scintillation placed less stringent constraints on source size, and thus was the interpretation they favored.
To evaluate whether the order unity variation we observed in Arp~220 can be explained as a result of interstellar scintillation within the Milky Way, we follow \citet{Darling2002b}, and use the results of \citet{Walker1998}.

At the galactic latitude of Arp~220 (b = 53.0$^\circ$), the scattering strength is unity at a frequency $\sim 8$~GHz, and the angular size above which a source cannot be considered a point source is $\sim 4$~$\mu$as \citep{Walker2001}.
In the strong scattering regime, the interstellar scattering disk along the line-of-sight to Arp~220 is then $\sim 0.13$~mas. 
To produce order unity variation, as we observed, the source size must be smaller than this scattering disk.
Variation produced by interstellar scintillation thus requires the physical size of varying masing spots to be $\lesssim$0.05~pc. 
As individual spots in E1, which have sizes 1--2~pc \citep{Lonsdale1998}, experienced order unity variations at some velocities, variation from scintillation would require significant substructure within the spots. 
The alternative explanation is changes within the masing clouds over the 14 years that elapsed between observations. 
With such a long period between observations, intrinsic variation is a much weaker constraint on source size/structure.
Thus the evidence suggests significant intrinsic variation in the compact masing regions of Arp~220.

\section{Conclusions} \label{sec:h_conclusion}
We detected magnetic fields associated with three masing clouds in Arp~220.
The measured strengths are in reasonably good agreement with previous single dish observations, though single dish observations of Zeeman splitting in OHMs may moderately underestimate the strength of magnetic fields.
This agreement supports past interpretation of the magnetic fields observed in OHMs using single dish radio telescopes.
The measured field strengths, and the sizes of clouds with which they are associated, are consistent with magnetic fields being dynamically important within the clouds.
Comparison of our results with previous total intensity VLBI observations reveals variability in the maser lines, in some cases order unity variations.
While this variability could either be intrinsic or due to interstellar scintillation, intrinsic variation appears to be the more likely cause in Arp~220, given the magnitude and timescale of observed variation. \\

\section{Acknowledgments}
We are very grateful to Vincent Fish for advice on reducing VLBI polarization data, and to Hans-Rainer Kl\"{o}ckner for his online reference for getting started with AIPS.
We thank Athol Kemball and Jon Romney for advice on VLBI spectropolarimetry experiments, Emmanuel Momjian, and Cormac Reynolds for help preparing observing scripts, and Tapasi Ghosh, Frank Ghigo, Jules Harnett, and Craig Walker for assistance observing.
J. M. received support from a NSF Graduate Research Fellowship and from a UC Berkeley Dissertation Year Fellowship.
This research used NASA's Astrophysics Data System Bibliographic Services, the SIMBAD database, operated at CDS, Strasbourg, France, the NASA/IPAC Extragalactic Database (NED), which is operated by the Jet Propulsion Laboratory, California Institute of Technology, under contract with the NASA, and APLpy, an open-source plotting package for Python.

\bibliographystyle{mn2ealt}
\bibliography{ms,extra}

\begin{thebibliography}{41}
\expandafter\ifx\csname natexlab\endcsname\relax\def\natexlab#1{#1}\fi

\bibitem[{Baan, Wood \& Haschick(1982)Baan, Wood, \& Haschick}]{Baan1982}
Baan W.~A., Wood P. A.~D., Haschick A.~D., 1982, \apj, 260, L49

\bibitem[{Beck(2012)}]{Beck2012}
Beck R., 2012, Space Science Reviews, 166, 215

\bibitem[{{Beck} \& {Krause}(2005)}]{Beck2005}
{Beck} R., {Krause} M., 2005, Astronomische Nachrichten, 326, 414

\bibitem[{Boulares \& Cox(1990)}]{Boulares1990}
Boulares A., Cox D.~P., 1990, \apj, 365, 544

\bibitem[{{Caswell}, {Green} \& {Phillips}(2013){Caswell}, {Green}, \&
  {Phillips}}]{Caswell2013}
{Caswell} J.~L., {Green} J.~A., {Phillips} C.~J., 2013, \mnras, 431, 1180

\bibitem[{{Caswell}, {Green} \& {Phillips}(2014){Caswell}, {Green}, \&
  {Phillips}}]{Caswell2014}
{Caswell} J.~L., {Green} J.~A., {Phillips} C.~J., 2014, \mnras, 439, 1680

\bibitem[{Condon {et~al}\mbox{.}(1991)Condon, Huang, Yin, \&
  Thuan}]{Condon1991}
Condon J.~J., Huang Z.-P., Yin Q.~F., Thuan T.~X., 1991, \apj, 378, 65

\bibitem[{Croston {et~al}\mbox{.}(2003)Croston, Hardcastle, Birkinshaw, \&
  Worrall}]{Croston2003}
Croston J.~H., Hardcastle M.~J., Birkinshaw M., Worrall D.~M., 2003, \mnras,
  346, 1041

\bibitem[{Crutcher {et~al}\mbox{.}(2010)Crutcher, Wandelt, Heiles, Falgarone,
  \& Troland}]{Crutcher2010}
Crutcher R.~M., Wandelt B., Heiles C., Falgarone E., Troland T.~H., 2010, \apj,
  725, 466

\bibitem[{Darling \& Giovanelli(2002)}]{Darling2002b}
Darling J., Giovanelli R., 2002, \apjl, 569, L87

\bibitem[{Diamond {et~al}\mbox{.}(1989)Diamond, Norris, Baan, \&
  Booth}]{Diamond1989}
Diamond P.~J., Norris R.~P., Baan W.~A., Booth R.~S., 1989, \apj, 340, L49

\bibitem[{{Elitzur}(1992)}]{Elitzur1992book}
{Elitzur} M., ed., 1992, Astrophysics and Space Science Library, Vol. 170,
  {Astronomical masers}

\bibitem[{Fish {et~al}\mbox{.}(2005)Fish, Reid, Argon, \& Zheng}]{Fish2005}
Fish V.~L., Reid M.~J., Argon A.~L., Zheng X., 2005, \apjs, 160, 220

\bibitem[{Graham {et~al}\mbox{.}(1990)Graham, Carico, Matthews, Neugebauer,
  Soifer, \& Wilson}]{Graham1990}
Graham J.~R., Carico D.~P., Matthews K., Neugebauer G., Soifer B.~T., Wilson
  T.~D., 1990, \apj, 354, L5

\bibitem[{Greisen(2003)}]{Greisen2003}
Greisen E.~W., 2003, {AIPS, the VLA, and the VLBA}, Vol. 285. Springer
  Netherlands, Dordrecht, p. 109

\bibitem[{Hardcastle, Birkinshaw \& Worrall(1998)Hardcastle, Birkinshaw, \&
  Worrall}]{Hardcastle1998}
Hardcastle M.~J., Birkinshaw M., Worrall D.~M., 1998, \mnras, 294, 615

\bibitem[{{Heiles} \& {Crutcher}(2005)}]{Heiles2005}
{Heiles} C., {Crutcher} R., 2005, in Lecture Notes in Physics, Berlin Springer
  Verlag, Vol. 664, Cosmic Magnetic Fields, {Wielebinski} R., {Beck} R., eds.,
  p. 137

\bibitem[{Heiles {et~al}\mbox{.}(1993)Heiles, Goodman, McKee, \&
  Zweibel}]{Heiles1993}
Heiles C., Goodman A.~A., McKee C.~F., Zweibel E.~G., 1993, In: Protostars and
  planets III (A93-42937 17-90), 279

\bibitem[{Kazes \& Baan(1991)}]{Kazes1991}
Kazes I., Baan W.~A., 1991, \aap, 248, L15

\bibitem[{Kus {et~al}\mbox{.}(1988)Kus, Marecki, Neff, van Ardenne, \&
  Wilkinson}]{Kus1988}
Kus A.~J., Marecki A., Neff S., van Ardenne A., Wilkinson P.~N., 1988, IN: The
  impact of VLBI on astrophysics and geophysics; Proceedings of the 129th IAU
  Symposium

\bibitem[{Liszt(2002)}]{Liszt2002a}
Liszt H., 2002, Publications of the Astronomical Society of the Pacific, 114,
  1087

\bibitem[{Lo(2005)}]{Lo2005}
Lo K., 2005, \araa, 43, 625

\bibitem[{Lockett \& Elitzur(2008)}]{Lockett2008}
Lockett P., Elitzur M., 2008, \apj, 677, 985

\bibitem[{Lonsdale, Diamond \& Smith(1994)Lonsdale, Diamond, \&
  Smith}]{Lonsdale1994a}
Lonsdale C. C. J. C.~J., Diamond P.~J., Smith H. H.~E., 1994, Nature, 370, 117

\bibitem[{Lonsdale {et~al}\mbox{.}(2007)Lonsdale, de~Kleer, Diamond, Thrall,
  Lonsdale, \& Smith}]{Lonsdale2007}
Lonsdale C.~J., de~Kleer K.~R., Diamond P.~J., Thrall H., Lonsdale C.~J., Smith
  H.~E., 2007, Proceedings of the International Astronomical Union, 3, 432

\bibitem[{Lonsdale {et~al}\mbox{.}(1998)Lonsdale, Diamond, Smith, \&
  Lonsdale}]{Lonsdale1998}
Lonsdale C.~J., Diamond P.~J., Smith H.~E., Lonsdale C.~J., 1998, \apj, 493,
  L13

\bibitem[{McBride \& Heiles(2013)}]{McBride2013}
McBride J., Heiles C., 2013, \apj, 763, 8

\bibitem[{McBride \& McCourt(2014)}]{McBride2014a}
McBride J., McCourt M., 2014, \mnras, 442, 838

\bibitem[{McBride {et~al}\mbox{.}(2014)McBride, Quataert, Heiles, \&
  Bauermeister}]{McBride2014}
McBride J., Quataert E., Heiles C., Bauermeister A., 2014, \apj, 780, 182

\bibitem[{Parra {et~al}\mbox{.}(2007)Parra, Conway, Diamond, Thrall, Lonsdale,
  Lonsdale, \& Smith}]{Parra2007}
Parra R., Conway J.~E., Diamond P.~J., Thrall H., Lonsdale C.~J., Lonsdale
  C.~J., Smith H.~E., 2007, \apj, 659, 314

\bibitem[{Parra, Conway \& Elitzur(2005)Parra, Conway, \& Elitzur}]{Parra2005}
Parra R., Conway J.~E., Elitzur M., 2005, \apss, 295, 325

\bibitem[{Robishaw, Quataert \& Heiles(2008)Robishaw, Quataert, \&
  Heiles}]{Robishaw2008}
Robishaw T., Quataert E., Heiles C., 2008, \apj, 680, 981

\bibitem[{Rovilos {et~al}\mbox{.}(2003)Rovilos, Diamond, Lonsdale, Lonsdale, \&
  Smith}]{Rovilos2003}
Rovilos E., Diamond P.~J., Lonsdale C.~J., Lonsdale C.~J., Smith H.~E., 2003,
  \mnras, 342, 373

\bibitem[{Sarma {et~al}\mbox{.}(2005)Sarma, Momjian, Troland, \&
  Crutcher}]{Sarma2005}
Sarma A.~P., Momjian E., Troland T.~H., Crutcher R.~M., 2005, \aj, 130, 2566

\bibitem[{Smith \& Lonsdale(1998)}]{Smith1998}
Smith H., Lonsdale C., 1998, \apj, 492, 137

\bibitem[{{Stahler} \& {Palla}(2005)}]{Stahler2005}
{Stahler} S.~W., {Palla} F., 2005, {The Formation of Stars}. Wiley-VCH

\bibitem[{Thompson {et~al}\mbox{.}(2006)Thompson, Quataert, Waxman, Murray, \&
  Martin}]{Thompson2006}
Thompson T.~A., Quataert E., Waxman E., Murray N., Martin C.~L., 2006, \apj,
  645, 186

\bibitem[{Walker(1998)}]{Walker1998}
Walker M.~A., 1998, \mnras, 294, 307

\bibitem[{Walker(2001)}]{Walker2001}
Walker M.~A., 2001, \mnras, 321, 176

\bibitem[{Willett {et~al}\mbox{.}(2011{\natexlab{a}})Willett, Darling, Spoon,
  Charmandaris, \& Armus}]{Willett2011a}
Willett K.~W., Darling J., Spoon H. W.~W., Charmandaris V., Armus L.,
  2011{\natexlab{a}}, \apjs, 193, 18

\bibitem[{Willett {et~al}\mbox{.}(2011{\natexlab{b}})Willett, Darling, Spoon,
  Charmandaris, \& Armus}]{Willett2011}
Willett K.~W., Darling J., Spoon H. W.~W., Charmandaris V., Armus L.,
  2011{\natexlab{b}}, \apj, 730, 56

\end{thebibliography}

\end{document}